\documentclass{article}

\usepackage{arxiv}

\usepackage[utf8]{inputenc} 
\usepackage[T1]{fontenc}    
\usepackage{hyperref}
\hypersetup{colorlinks,allcolors=black}
\usepackage{url}            
\usepackage{booktabs}       
\usepackage{amsfonts}       
\usepackage{amsmath}
\usepackage{amssymb}
\usepackage{bm}             
\usepackage{nicefrac}       
\usepackage{doi}
\usepackage{graphicx}       

\usepackage[
backend=biber,
style=nature, 
url = false,
isbn = false,
]{biblatex}
\usepackage{setspace}

\usepackage{authblk}    

\usepackage[
figurename=Figure, 
labelfont=bf, 
labelsep=space,
font=onehalfspacing
]{caption}

\usepackage{algorithm}
\usepackage[noend]{algpseudocode}

\usepackage{siunitx} 	

\usepackage{array}
\usepackage{multirow}
\usepackage{mathtools}
\usepackage{braket}
\usepackage{bbold}

\author[1]{Mathieu Suter}
\author[2]{Jens P. Metzger}
\author[1]{Andreas Port}
\author[2]{Christoph R. M\"{u}ller}
\author[1,*]{Klaas P. Pruessmann}

\affil[1]{Institute for Biomedical Engineering, ETH Zurich and University of Zurich, 8092 Zurich, Switzerland}

\affil[2]{Department of Mechanical and Process Engineering, ETH Zurich, 8092 Zurich, Switzerland}

\affil[*]{Corresponding author: pruessmann@biomed.ee.ethz.ch}

\begin{document}

\title{Magnetic Resonance Particle Tracking}

\date{March 28, 2025}
\maketitle

\doublespacing

\begin{abstract}
Granular materials such as gravel, cereals, or pellets, are ubiquitous in nature, daily life, and industry. While sharing some characteristics with gases, liquids, and solids, granular matter exhibits a wealth of phenomena that defy these analogies and are yet to be fully understood. Advancing granular physics requires experimental observation at the level of individual particles and over a wide range of dynamics. Specifically, it calls for the ability to track large numbers of particles simultaneously, in three dimensions (3D), and with high spatial and temporal resolution. Here, we introduce magnetic resonance particle tracking (MRPT) and show it to achieve such recording with resolution on the scale of micrometers and milliseconds. Enabled by MRPT, we report the direct study of granular glassy dynamics in 3D. Tracking a vibrated granular system over six temporal orders of magnitude revealed dynamical heterogeneities, two-step relaxation, and structural memory closely akin to 3D glass formation in supercooled liquids and colloids. These findings illustrate broad prospective utility of MRPT in advancing the exploration, theory, and numerical models of granular matter.
\end{abstract}



\section{Main}

Granular matter is athermal and dissipative in nature and exhibits a range of peculiar phenomena such as avalanching, segregation, jamming, and glass formation \cite{RN1,RN2,RN3,RN4,RN5}. Many aspects of granular behavior are yet to be fully explored and lack a fundamental understanding. Experimental insight to these ends must reach beyond the macroscale due to an inherent absence of scale separation \cite{RN6}. Granular dynamics thus need to be studied at the level of individual particles, calling for means of particle tracking.

A range of tracking techniques have been reported \cite{RN7}. For 2D and quasi-2D systems, one effective way of capturing individual particle motion is by a high-speed camera \cite{RN8,RN9}. Probing granular matter in 3D requires physics to which the system in question is largely transparent. For instance, X-ray tomography has been used to locate particles in quasi-static conditions or subject to step motion interleaved with scanning at rest \cite{RN10,RN11,RN12}. Dynamic granular systems such as fluidized beds or rotating drums have been investigated by positron emission particle tracking (PEPT), which uses radioactively labelled tracer particles \cite{RN13,RN14}. However, due to the underlying physics, PEPT has limited capability to distinguish tracers close to each other and is hence typically performed with a single one.

In this work, we deploy a different physical mechanism, nuclear magnetic resonance (NMR), to advance 3D particle tracking. Magnetic resonance is the basis of MRI, an imaging modality common in medicine and the life sciences that has also been used to study granular systems \cite{RN15,RN16,RN17,RN18,RN19,RN20}. However, offering resolution on the order of seconds and millimeters, 3D MRI is unsuited for capturing particle dynamics. Instead, we use NMR in a fundamentally different way that exploits the discrete nature of granular matter in conjunction with signal modulation at critical wavelength. As we show, magnetic resonance particle tracking (MRPT) resolves granular dynamics on the scale of milliseconds and micrometers, simultaneously for large numbers of particles, and with preservation of particle identity.

One frontier, on which these capabilities make a key difference, is the study of glassy behavior in granular systems. Understanding glass is one of the most challenging open problems in condensed matter physics and materials science \cite{RN5,RN21,RN22}. In thermal systems, a common way to form a glass is by supercooling, causing a stark increase in viscosity and quenching the system into an amorphous solid \cite{RN21}. Similarly, when externally driven granular matter is increasingly compressed, it undergoes a slowdown of particle dynamics, eventually freezing into a disordered jammed state \cite{RN4}. This correspondence is remarkable in that granular matter is athermal and dissipative. The underlying microscopic mechanisms are reflected in dynamics and correlations at the particle level. So far, tracking of granular glassy dynamics has been limited to 2D systems \cite{RN8,RN9,RN23}. Here, we report the study of such dynamics in 3D, using MRPT. We find them to be closely akin to those of supercooled liquids and colloids, albeit on vastly different spatiotemporal scales.

\subsection{Principle}

\begin{figure}
    \centering
    \includegraphics[width=1\linewidth]{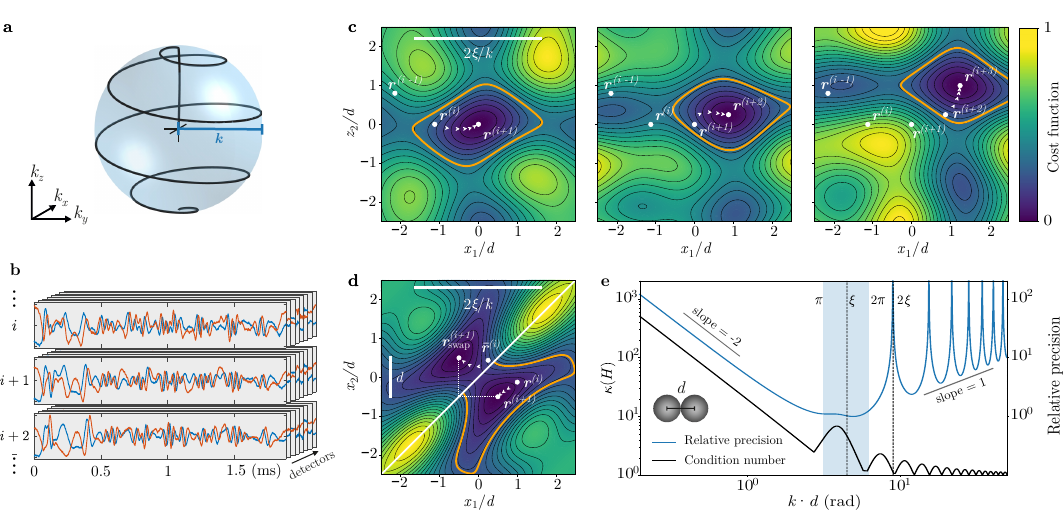}
    \caption{\textbf{| Fundamentals of MRPT.}\em{ \textbf{a}, k-space trajectory (black) on a sphere of radius $k$, reflecting spatial signal modulation.\textbf{ b}, Time series of complex-valued signals pertaining to successive frames $i$, $i+1$, $i+2$, acquired with multiple detectors. \textbf{c}, Visualization of the tracking process. The cost function is shown for the subspace spanned by the \textit{x} coordinate of one tracer ($x_1$) and the $z$ coordinate of another tracer ($z_2$). Local optimization for frame $i+1$ is initialized with the preceding configuration ${\boldsymbol{r}}^{\left(i\right)}$ and converges to the current configuration ${\boldsymbol{r}}^{\left(i+1\right)}$ (white arrowheads). This procedure is performed recursively. The orange contour line is a visual guide for the neighborhood of safe convergence. The scale of the features of the cost function is $2\xi /k$, where $\xi $ is the first zero of the spherical Bessel function $j_1$. \textbf{d}, Cost function of two tracers in contact along the $x$ coordinate. The two global minima correspond to the actual tracer configuration and its counterpart with tracers swapped. To avoid swapping, the preceding configuration should not be across the diagonal (white line). \textbf{e}, Relative precision (blue) of tracer coordinates as a function of $k$ in the case of two touching tracers. Vertical asymptotes indicate full signal cancelation by dephasing. The condition number of the cost-function Hessian at the global minimum is shown in black. Near-optimal precision is achieved in a critical range $2\pi \gtrsim kd\gtrsim \pi$ (blue band). }}
    \label{fig:1}
\end{figure}

For MRPT, particles with NMR-active cores are added to a granular system under study as tracers. Tracking is then based on rapid succession of NMR measurements at intervals $\Delta$t of few milliseconds. Magnetic gradient fields are used to sensitize the NMR signals to particle position by spatial modulation with wave number $k$. The direction of modulation is varied continuously, corresponding to a trajectory embedded in a shell of radius $k$ in vectorial 3D k-space (Fig.~\ref{fig:1}a). Concurrent sensing with an array of NMR detectors yields sets of signals (Fig.~\ref{fig:1}b). Assuming spherical tracer cores of diameter $d$ and corresponding volume $V$, the signals are modeled as 
\begin{equation*}
    f_{c,n}\left({\boldsymbol{r}}_1,\dots,{\boldsymbol{r}}_{N_p},s_{c,1}\dots ,s_{c,N_p}\right)={\mathcal{F}}_B\left({|\boldsymbol{k}}_n|\frac{d}{2}\right)V\sum^{N_p}_{p=1}{s_{c,p} e^{-i{\boldsymbol{k}}_n\cdot{\boldsymbol{r}}_p}}
\end{equation*}
where $p$ counts tracers, $c$ counts detectors, $s_{c,p}$ denotes detector sensitivity at position ${\boldsymbol{r}}_p$, $n$ counts signal samples taken at k-space positions ${\boldsymbol{k}}_n$ and ${\mathcal{F}}_B$ is the Fourier transform of the unit ball, reflecting within-tracer dephasing.

Given a measurement $\{y_{n,c}\}$ of a certain particle configuration, we reconstruct tracer positions and detector sensitivities by finding the global minimum of the cost function $\mathcal{L}=\sum_{c,n}{{\left|y_{c,n}-f_{c,n}\left({\boldsymbol{r}}_1,\dots,{\boldsymbol{r}}_{N_p},s_{c,1},\dots,s_{c,N_p}\right)\right|}^2}$, where ${\boldsymbol{r}}_p$ and $s_{c,p}$ are unknowns. On its own, this optimization is generally intractable due to high dimensionality and non-convexity of the cost function. We solve this problem by exploiting the continuity of tracer motion, leading to a recursive procedure that tracks the global minimum of $\mathcal{L}$ along with the tracers.

This process is illustrated in Fig.~\ref{fig:1}c. In each successive frame, a position update is obtained by local optimization initialized by the configuration found in the preceding frame. To keep track, for each local optimization the respective previous configuration needs to be inside the basin of attraction of the global minimum. The size of the basin is inversely proportional to the k-space radius $k$ and the scale of features$\ $in $\mathcal{L}$ is about $2\xi/k$, where $\xi$ is the first zero of the spherical Bessel function $j_1$. Thus, for the reconstruction to track the global minimum, the maximum tracer displacement per $\Delta t$ must not exceed a fraction of that length. A sufficient condition for safe tracking is $2k{\max}_{p}|\Delta{\boldsymbol{r}}_p|<\xi$, where $\Delta{\boldsymbol{r}}_p$ denotes the displacement of tracer $p$.

A second condition arises from symmetry. For physically identical tracers, the signal model and hence the cost function are even under permutation of the tracers. This leads to the existence of alternate global minima, posing the risk of swapping tracers during reconstruction. Alternate minima come closest when tracers are directly adjacent or collide as shown in Fig.~\ref{fig:1}d and Supplementary Video 1. A sufficient condition for preserving tracer identity is for the displacement of tracers per frame to be less than the tracer radius ($|\Delta{\boldsymbol{r}}_p|<d/2$).

Tracking precision is governed by propagation of thermal, Gaussian detection noise. Given the Hessian matrix $H$of the cost function $\mathcal{L}$ at the global minimum, a global metric capturing noise in resulting particle coordinates is $\sqrt{\mathrm{Tr}\left[H^{-1}\right]}$, and, as one varies $k$, it behaves as $\frac{\sqrt{\kappa\left(H\right)}}{k{|\mathcal{F}}_B\left(kd/2\right)|V}$ where $\kappa\left(H\right)$ is the condition number of $H$. Fig.~\ref{fig:1}e shows relative precision alongside $\kappa\left(H\right)$ as a function of $k$ in the limiting case where two tracers are touching (see Extended Data Fig.~\ref{fig:1} for the case where they are distant). Interestingly, the condition number curve is reminiscent of diffraction behavior. Specifically, it resembles the Mie solution of a plane wave scattered by a sphere, such as the radar cross section of a sphere of diameter $2d$ \cite{RN24}. Indeed, we can observe a Rayleigh region for $kd\ll 1$ where the condition number behaves as a power law, an intermediate oscillatory Mie region, and an optical region for $kd\gg 1$ where the wavelength is much smaller than the particle diameter. For small $k$, small curvature of the cost function and bad conditioning lead to poor precision. When $k$ is too large, tracer dephasing degrades signal yield, equally at the expense of precision. Near-optimal precision is obtained with $k$ such that $2\pi \gtrsim kd \gtrsim\pi $. Note that this critical range coincides with the start of the Mie scattering region (first peak) when the probing wavelength is of the order of the particle size or smaller.

Importantly, these three demands, regarding safety of tracking, preservation of identity, and optimality of precision, can be reconciled. Given tracers of diameter $d$ and expected system dynamics, we limit the maximum displacement to $d/2$ by choosing a suitable $\Delta t$. Independently, $k$ is selected such that $kd\approx \xi $, minimizing noise propagation. Based on these choices, the tracking condition is met automatically.

\subsection{Tracking performance}

\begin{figure}
    \centering
    \includegraphics[width=0.8\linewidth]{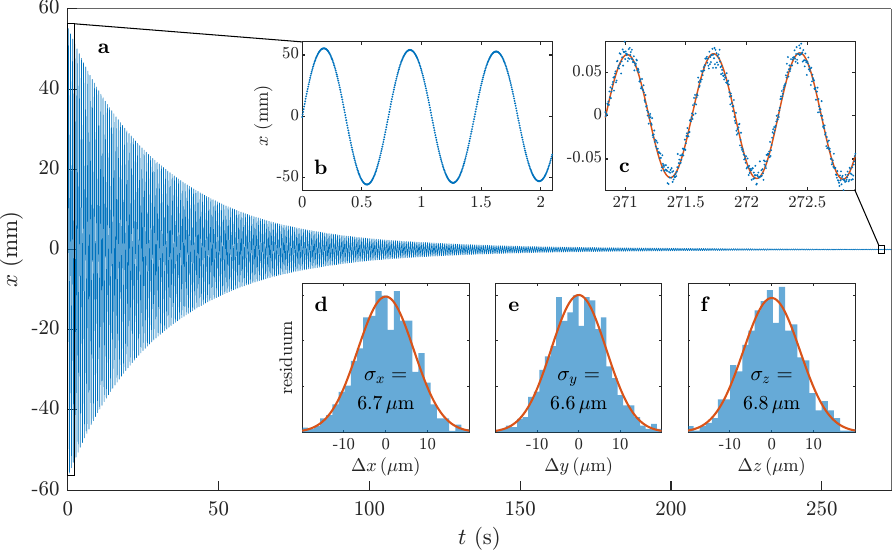}
    \caption{\textbf{| Tracking a pendulum.} \em{\textbf{a}, Time series of positions of a tracer oscillating in the $x$-$z$ plane, where \textit{z} is aligned with gravity. The $x$ component is displayed. \textbf{b},\textbf{ }Early oscillations with a range of motion of 10~cm. \textbf{c}, Later oscillations, where the range of motion has decayed to about 100 \textmu{}m. A damped harmonic oscillator model was fitted to the data (orange line). \textbf{d-f}, Distribution of the residua of the fit in \textbf{c} for each coordinate (blue bars). Each distribution is fitted by a Gaussian (orange line) with standard deviation $\sigma $ indicating spatial precision.}}
    \label{fig:2}
\end{figure}

To assess the precision of MRPT, we track a pendulum oscillating in a vertical plane over a duration of 280~s, at a temporal resolution of $\Delta t$~= 4.3~ms (Fig.~\ref{fig:2}a). For small oscillations, the system behaves like a damped harmonic oscillator. This model is fitted to the late part of the trajectory (Fig.~\ref{fig:2}c) where the amplitude of motion is less than 100~\textmu{}m. The histograms of the residua (Fig.~\ref{fig:2}d-f) indicate that MRPT is capable of tracking motion at single-digit micrometer precision uniformly in all directions, capturing particle dynamics over five orders of magnitude in space and time. 

\begin{figure}
    \centering
    \includegraphics[width=1\linewidth]{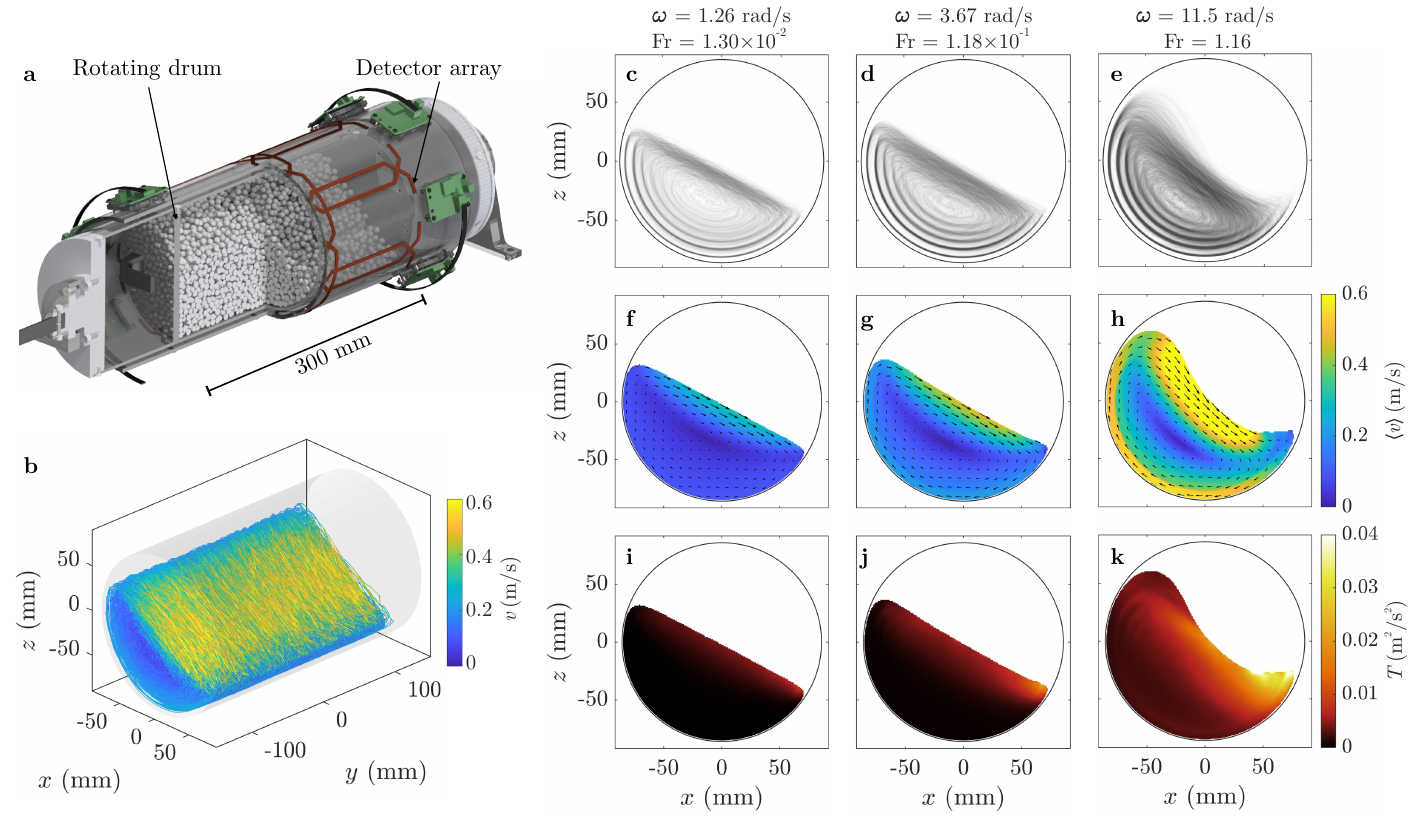}
    \caption{\textbf{| Granular dynamics in a rotating drum. }\em{\textbf{a}, Setup. \textbf{b}, 3D trajectories of 10 tracers obtained during 25~min of rotation at 3.67~rad/s. The color coding indicates the velocity magnitude. \textbf{c}-\textbf{e}, Axial projections of trajectories obtained at three different rotation rates $\omega $ and corresponding Froude numbers $\mathrm{Fr}$ (only 4.3~min shown for better visibility). \textbf{f-h}, Mean velocity field, reflecting the quasi-solid region, the active layer, and their transitions. \textbf{i-k}, High spatiotemporal resolution permits detailed mapping of the granular temperature.}}
    \label{fig:3}
\end{figure}

The ability to track multiple particles and more complex motion is demonstrated in a rotating drum (Fig.~\ref{fig:3}a). Rotating drums exhibit a broad range of granular dynamics with coexisting granular phases, namely a quasi-solid region superimposed by liquid- and gas-like regions in a collision-dominated active layer \cite{RN25,RN26,RN27,RN28}. We track 10 tracers of 8~mm diameter simultaneously, rendering the tracking problem high-dimensional with 30 and 120 unknowns, respectively, for particle coordinates and detector sensitivities. The resulting complex structure and dynamics of the cost function are illustrated in Supplementary Video 2.

Particle trajectories obtained by 25~min of tracking (Fig.~\ref{fig:3}b) show the slow quasi-solid region and fast active layer in the rolling regime. The transition from rolling to cataracting, including the appearance of ballistic motion, is illustrated by axial projections of trajectories obtained with increasing rotation rate $\omega$ (Fig.~\ref{fig:3}c-e). These show partial crystallization in the quasi-solid region, whereas trajectories in the active layer are less ordered. Eulerian mean velocity fields are derived from tracer velocities (Fig.~\ref{fig:3}f-h), illustrating the formation of the active layer. Fluctuations about this mean constitute granular temperature, a key quantity governing mass, momentum and energy transport in rapid granular flows \cite{RN29,RN30}. High-resolution tracking enabled detailed mapping of granular temperature (Fig.~\ref{fig:3}i-k). It is highest where the active layer collides with the wall while lower temperatures are seen in the quasi-solid and ballistic regions (Fig.~\ref{fig:3}k). Detection noise introduces fluctuation in tracked positions, which appears as a spurious temperature baseline. However, the noise-induced baseline amounts to only 1\% of the peak granular temperature, reflecting high tracking precision.

Animated displays of the different rotating-drum regimes are provided in Supplementary Videos 3-4, showing the dynamics of avalanching, rolling and cataracting with 20 tracers simultaneously tracked.

\subsection{Granular glassy dynamics}

\begin{figure}
    \centering
    \includegraphics[width=1\linewidth]{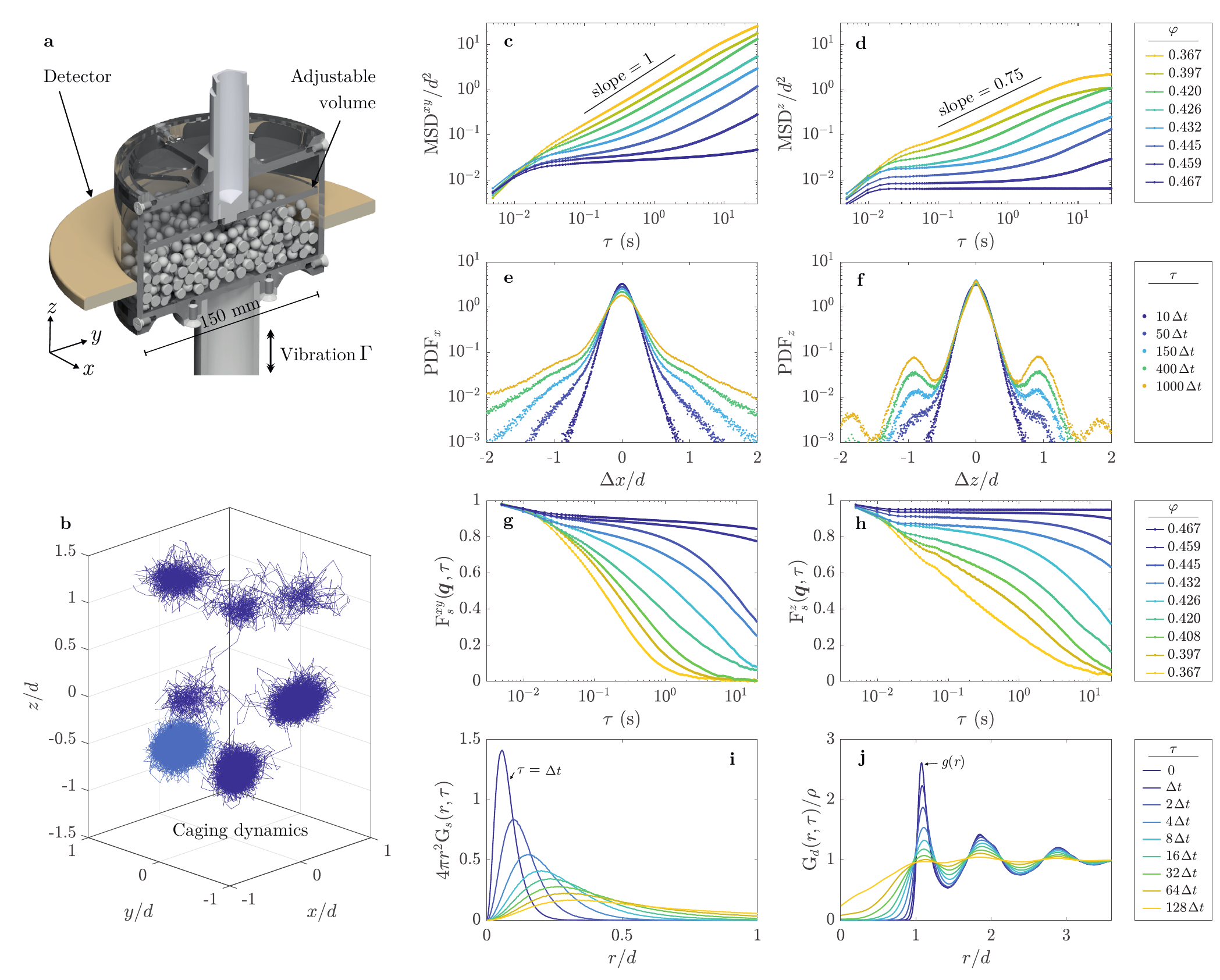}
    \caption{\textbf{| Glassy dynamics in a 3D vibrated granular bed.}\em{ \textbf{a}, Cylindrical container filled with 1000 spherical particles ($d=8\,\mathrm{mm}$) and 6 mechanically identical tracers. The system is vertically vibrated at 30~Hz with dimensionless acceleration $\Gamma \approx6.5$ and tracked at a temporal resolution of $\Delta t=4.9\,\mathrm{ms}$. \textbf{b}, Trajectories of two initially adjacent tracers for 100~s at volume fraction $\varphi=0.432$. \textbf{c-d}, MSD of the tracers in the horizontal plane (\textbf{c}) and the vertical direction (\textbf{d)} for varying $\varphi $. \textbf{e-f}, PDF of horizontal (\textbf{e}) and vertical (\textbf{f}) displacements at different times at $\varphi =0.432$. \textbf{g-h}, Self-part of the ISF for $|\boldsymbol{q}|=4/d$ and varying $\varphi $. \textbf{i-j}, Self-part (\textbf{i}) and distinct part (\textbf{j}) of the van Hove function as a function of the 3D displacement $r$ at different times, $\varphi =0.408$. The distinct part is normalized by the density $\rho$ and for $\tau=0$, equals the pair correlation function. Note that the curves for the distinct part have been smoothed as detailed in the Methods section and Supplementary Information.}}
    \label{fig:4}
\end{figure}

Deploying MRPT, 3D glassy dynamics have been studied in a continuously vibrated granular bed (Fig.~\ref{fig:4}a). High spatiotemporal resolution has captured 3D caging dynamics, which are a characteristic feature of glass formation (Fig.~\ref{fig:4}b) (see Supplementary Videos 5-6 for animated displays of tracking results and underlying cost function dynamics). With the two shown tracers initially close, one remains close to its initial position whereas the other performs successive jumps, highlighting the presence of spatially heterogeneous dynamics \cite{RN31,RN32}. Next, the volume fraction $\varphi $ was varied and 6 tracers were tracked for one hour at a time. We first analyze the particle trajectories in terms of their mean squared displacement $\mathrm{MSD}\left(\tau \right)=\left\langle {\left|{\boldsymbol{r}}_p\left(\tau \right)-{\boldsymbol{r}}_p\left(0\right)\right|}^2\right\rangle$, where ${\boldsymbol{r}}_p\left(\tau \right)$ is the position of tracer $p$ at time $\tau$ and the brackets indicate an ensemble average. Fig.~\ref{fig:4}c shows the MSD in the horizontal ($x$-$y$) plane. At low $\varphi $ and short $\tau $, the plot reflects a transition from ballistic motion to the diffusive regime, which is characterized by a slope of 1. As $\varphi$ increases, the MSD plateaus due to tracers being locked in cages. As $\tau$ increases, the tracers jump from cage to cage, amounting to slow, long-range diffusion indicative of a two-step relaxation process \cite{RN5}. The same behavior is apparent in the probability density function (PDF) of the displacement in $x$ (Fig.~\ref{fig:4}e). The distribution deviates from Gaussian behavior by exhibiting exponential tails which is a further signature of dynamical heterogeneities \cite{RN31}. In the vertical direction, significant plateauing is again observed in the MSD (Fig.~\ref{fig:4}d). However, the long-time relaxation involves anomalous diffusion with a slope of approximately 0.75. Peaks in the PDF (Fig.~\ref{fig:4}f) show that particles preferentially reside in layers, reflecting partial crystallization in the vertical direction promoted by monodispersity of the particles.

The time scales of relaxation become more apparent studying the intermediate scattering function \cite{RN5,RN33}. Its self-part $F_s$ is given by
\begin{equation*}
    F_s\left(\boldsymbol{q},\tau \right)=\left\langle {\exp \left[-i\boldsymbol{q}\cdot ({\boldsymbol{r}}_p\left(\tau \right)-{\boldsymbol{r}}_p\left(0\right))\right]\ }\right\rangle ,
\end{equation*}
where $\boldsymbol{q}$ is the wave vector. The ISF for increasing $\varphi $ is shown in Fig.~\ref{fig:4}g-h. The two-step relaxation process is visible in the horizontal and vertical direction. The relaxation times increase with $\varphi $ and are longer in the vertical direction. 

The ability to track multiple particles over long time while preserving identity enables us to accurately capture their spatiotemporal correlations, which characterize structure in particulate systems. Fig.~\ref{fig:4}i-j show the correlation found in the vibrated bed in terms of the van Hove function \cite{RN5,RN33} $G(r,\tau )$:
\begin{equation*}
    {G= G}_s\left(r,\tau \right)+ G_d\left(r,\tau \right)=\left\langle \delta \left(r-\left|{\boldsymbol{r}}_p\left(\tau \right)-{\boldsymbol{r}}_p\left(0\right)\right|\right)\right\rangle +{\left\langle \delta \left(r-\left|{\boldsymbol{r}}_p\left(\tau \right)-{\boldsymbol{r}}_q\left(0\right)\right|\right)\right\rangle }_{p\neq q}\,,
\end{equation*}

where $G_s$ and $G_d$ denote the self- and distinct parts of the function, respectively. Given a particle $p$ at the origin at time zero, $G_s$ and $G_d$ represent the probability of finding the same ($p$) or another ($q$) particle at a distance $r$ from the origin at time $\tau $. Fig.~\ref{fig:4}j shows $G_d$ normalized by density $\rho $. For $\tau =0$, it is equal to the pair correlation function and shows features typical of dense fluids \cite{RN33}. As time increases, the system gradually loses memory of the local order. However, the probability of replacing one particle with another at $r<d$ remains very low, even at the time scale of the early diffusion regime. This reflects the fact that less mobile particles dominate $G_d$ up to roughly the particle diameter while more mobile particles dominate otherwise. Dynamical heterogeneities cause the system to already appear diffusive while some memory of the local order still remains. Notably, the shapes and the evolution of $G_s$ and $G_d$ in the vibrated bed closely resemble those observed in colloids and thermal glass forming systems \cite{RN5,RN34,RN35,RN36}. The distortion of the second peak of $G_d$ is reminiscent of the peak splitting observed in liquids \cite{RN36}, colloids \cite{RN37} and other glass formers \cite{RN38}. 

In concert, these findings indicate that glassy behavior of 3D granular systems closely follows the same patterns as that of other glass forming systems even though they occur on vastly different spatial and temporal scales. They thus support the notion of universality of glass formation across thermal and driven, dissipative systems. 

\subsection{Outlook}

According to these results, MRPT achieves versatile particle tracking in opaque, dynamic 3D systems, reconciling high resolution with multiple-particle capability and preservation of particle identity. At current resolutions on the scale of milliseconds and micrometers, MRPT captures rapid dynamics on a fine length scale as highlighted by measurement of granular temperature and the observation of glassy dynamics down to the ballistic regime. Going forward, we expect MRPT to reach even higher resolution. Spatiotemporal resolution is governed primarily by the speed of spatial encoding and the signal-to-noise ratio (SNR) of the NMR signals. Speed-up of encoding will be achieved by boosting the strength and switching rate of the gradient fields involved \cite{RN39,RN40}. For greater sensitivity, one immediate approach is to increase the background field strength $B_0$, which will boost SNR and thus tracking precision in proportion to $B^{7/4}_0$ \cite{RN41}. Sensitivity is closely related also to tracer size as signal strength scales with the volume of NMR-active material. Higher SNR at higher field, or reduced demand in terms of precision, will thus enable the use of smaller tracers. Gains in both overall sensitivity and spatial encoding are also available from advancing array detection in terms of the number and density of detector elements \cite{RN42,RN43,RN44}. A promising perspective, in terms of both sensitivity and speed, is to implement MRPT on the basis of electron spin resonance \cite{RN45,RN46} rather than NMR, exploiting the substantially stronger magnetism and faster relaxation of electrons.

The principle of MRPT fundamentally supports any number of tracers. At this stage, simultaneous tracking has been demonstrated for up to 20 particles. Increasing the tracer count, the key consideration is numerical conditioning, which in turn depends on k-space coverage. Greater k-space velocity from stronger gradient fields or, alternatively, longer readouts are thus expected to enable yet greater particle counts. While the particles tracked in this work are spheres of uniform sizes, MRPT will equally apply to large sets of polydisperse and arbitrarily shaped tracers \cite{RN47,RN48}. 

MRPT of polydisperse particles will be instrumental particularly for furthering the study of granular glasses. In a vibrated bed with monodisperse particles, we have observed glassy behavior and partial crystallization, up to the hard-sphere freezing point \cite{RN49}. In polydisperse systems \cite{RN50}, 3D tracking shall enable direct observation of the glass transition and its microscopic mechanisms, which remain to be fully understood \cite{RN51,RN52,RN53,RN54}. Advances on this front shall also further elucidate the universality of glass formation across systems and scales.

Beyond glassy dynamics, measurement of correlation, enabled by multi-particle capability with preservation of particle identity, holds great promise also for the study of other granular phenomena such as jamming \cite{RN55,RN56}, nonlocal effects of granular flow \cite{RN57}, and memory effects \cite{RN10,RN58}. Further systems of particular interest include chute flows, gas-fluidized beds, and shear cells, which exhibit intriguing phenomena such as avalanching \cite{RN59}, segregation, and pattern formation \cite{RN60}, and appear in many industrial applications.

Throughout granular systems, particle tracking captures individual dynamics of their fundamental constituents. In doing so, it also reveals collective behavior and provides a key to understanding how it emerges. Enhanced tracking capability with MRPT thus holds promise for the study of granular systems at large, offering a generic means of exploring granular phenomena, testing theory, and advancing numerical models.

\onehalfspacing
\printbibliography

\doublespacing

\newpage
\section{Statements}






\subsection{ Acknowledgements}

 We thank J. Overweg, M. Weiger, F. Hennel, R. L\"{u}chinger, M. Engel and A. Penn for sharing MR system expertise; C. Schildknecht, N. Conzelmann, C. G\"{u}nthner and T. Schmid for helping with hardware; A. Penn, A. Silberstein Martinez and T. Noguera Valencia for insightful discussions; and S. Scherr for manufacturing the setups. J.P.M. was supported by the Swiss National Science Foundation under grant no. 200020\_182692. A.P. was supported by the Swiss National Science Foundation BRIDGE Discovery grant no. 181023.

\subsection{ Author contributions}

 M.S and K.P.P. devised the tracking principle and developed the method. M.S. performed the technical implementation. M.S., J.P.M. and K.P.P. performed experimental exploration and advancement. M.S., J.P.M., C.R.M. and K.P.P. designed the experiments. J.P.M. designed and built the mechanical setups. A.P. developed and built the detector array. J.P.M. performed the mechanical design of the detector array. M.S. and J.P.M. performed the experiments and the data analysis. M.S., J.P.M. and K.P.P. wrote the manuscript. M.S., J.P.M., C.R.M. and K.P.P. reviewed the manuscript. K.P.P. initiated and guided the project.



\subsection{Additional information}

 Supplementary Information (including Methods and Videos) is available for this paper.
 Correspondence and requests for materials should be addressed to K.P.P. at {pruessmann@biomed.ee.ethz.ch}. 

\end{document}